\documentclass[prd,twocolumn,showpacs,preprintnumbers,nofootinbib,superscriptaddress,eqsecnum]{revtex4}

\usepackage{amsmath,amssymb,amsthm,color}
\usepackage{graphicx}
\usepackage{dcolumn}

\usepackage{xcolor}
\colorlet{darkblue}{blue!70!black}
\usepackage[colorlinks=true, urlcolor=darkblue,linkcolor=darkblue,citecolor=darkblue]{hyperref}
\usepackage{comment}

\usepackage{color}

\newcommand {\beq}{\begin{eqnarray}}
\newcommand {\eeq}{\end{eqnarray}}
\newcommand\be{\begin{equation}}
\newcommand\ba{\begin{eqnarray}}
\newcommand\ea{\end{eqnarray}}
\newcommand\ee{\end{equation}}
\newcommand{\lb}{\left(}
\newcommand{\rb}{\right)}

\newcommand{\lsb}{\left[}
\newcommand{\rsb}{\right]}

\newcommand{\nn}{\nonumber}
\newcommand{\pd}{\partial}
\newcommand\SR{\, ^*\hspace{-0.07cm}R}
\newcommand\SF{\, ^*\hspace{-0.07cm}F}
\newcommand\aCS{\alpha_\mathrm{CS}}
\newcommand\bCS{\beta_\mathrm{CS}}
\newcommand{\mrm}{\mathrm}

\newcommand{\hhref}[1]{}

\newcommand{\scO}{\ensuremath{\mathcal{O}}}

\newcommand\al{{\alpha}}
\newcommand\ep{\epsilon}

\newcommand\vt{\vartheta}

\newcommand\de{{\ensuremath{{\delta}}}}
\newcommand\De{{\ensuremath{{\Delta}}}}

\newcommand\nab{{\nabla}}

\newcommand\ov{\over}
\newcommand\ha{{1 \ov 2}}

\def\le{\left}
\def\ri{\right}

\newcommand\p{\ensuremath{\partial}}

\newcommand\sL{{\ensuremath{{\mathcal L}}}}

\newcommand\sO{{\ensuremath{{\mathcal O}}}}

\begin{document}
\preprint{CALT 68-2857, IPMU13-0193, MIT-CTP/4501}

\title{Angular momentum generation by parity violation}

\author{Hong Liu}
\affiliation{Center for Theoretical Physics,
Massachusetts Institute of Technology,
Cambridge, Massachusetts 02139, USA}
\author{Hirosi Ooguri}
\affiliation{California Institute of Technology, 452-48, Pasadena, California 91125, USA}
\affiliation{Kavli Institute for the Physics and Mathematics of the Universe (WPI),
University of Tokyo, Kashiwa 277-8583, Japan}
\author{Bogdan Stoica}
\email{bstoica@theory.caltech.edu}
\affiliation{California Institute of Technology, 452-48, Pasadena, California 91125, USA}

\date{\today}

\begin{abstract}
We generalize our holographic derivation of spontaneous angular momentum generation in $2 + 1$ dimensions 
in several directions. We consider cases when a parity-violating perturbation responsible for the angular momentum
generation can be nonmarginal (while in our previous paper we restricted to a marginal perturbation), including all possible 
two-derivative interactions, with parity violations triggered both by gauge and gravitational Chern-Simons terms in the bulk.
We make only a minimal assumption about the bulk geometry that it is asymptotically AdS, respects the Poincar\'e symmetry in $2 + 1$ dimensions,
and has a horizon. In this generic setup, we find a remarkably concise and universal formula for the expectation value of the angular momentum
density, to all orders in the parity violating perturbation.

\end{abstract}

\maketitle


\section{Introduction}

The spontaneous generation of angular momentum and of an edge current are typical phenomena in parity-violating physics~(see, for example,~\cite{Volovik,stone,hoyos,sauls,Nicolis:2011ey}).  For a given interacting system, 
whether spontaneous generation of angular momentum does occur, and if yes, the precise value, are important dynamical questions for which a universal answer (applicable to generic parity-violating systems) does not appear to exist.  A famous example is helium 3-A, in which case there has been a long controversy about the value of its angular momentum (see e.g.~\cite{Volovik,Leggett}). The controversy highlights the importance of finding exactly solvable models, especially strongly interacting systems, through which one could extract generic lessons. Holographic systems are ideal laboratories for this purpose. 

In a previous paper~\cite{Liu:2012zm}, we initiated exploration of these phenomena in 
holographic systems.\footnote{See~\cite{anomalyone, anomalytwo, modulatedone, modulatedtwo, SaremiSon, Chen:2011fs, Jensen:2011xb, Chen:2012ti, qed1, qed2, qed3} for other discussions of parity-violating effects in holographic systems and in $(2+1)$-dimensional field theories.} There, for technical simplicity, we restricted to parity violation effected by turning on a marginal pseudoscalar operator, and considered only the Schwarzschild and Reissner-Nordstr\"om geometries. In this paper, we generalize the results to parity violation through a relevant scalar operator, and to general bulk black hole geometries.  

More explicitly, we consider a $(2+1)$-dimensional boundary field theory with a $U(1)$ global symmetry, which is described by classical gravity (together with various matter fields) in a four-dimensional, asymptotically anti--de Sitter spacetime~(AdS$_4$).   We consider two representative bulk mechanisms for parity violation, with a gravitational Chern-Simons interaction~\cite{CSgravity}
 \be \label{gcs}
\aCS \int \vartheta\ R \wedge R, 
\ee
or an axionic coupling~\cite{Wilczek:1987, Carroll:1989vb} 
\be  \label{axio}
\bCS
 \int \vartheta \ F \wedge F , 
\ee 
where $R$ is the Riemann curvature two-form, $F$ is field strength for the bulk gauge field $A_a$ dual to the $U(1)$ global current, and  $\vartheta$ is a pseudoscalar dual to a boundary relevant pseudoscalar operator $\scO$. 
$\aCS$ and $\bCS$ are some constants.

 The parity symmetry is broken explicitly if a source is turned on for $\scO$ corresponding to turning on a  
 non-normalizable mode for the pseudoscalar field $\vartheta$. Alternatively, the parity can be spontaneously 
 broken when $\scO$ develops an expectation value in which case 
 the bulk field $\vartheta$ is  normalizable.
In both situations if we put the system in a finite box (i.e., parity-violation terms are nonzero only inside the box), 
the spontaneous generation of angular momentum is always accompanied by an edge current.
We emphasize that the source or expectation value for $\scO$ is taken to be homogeneous along 
boundary directions. An angular momentum density is generated, despite the boundary quantum state and the corresponding bulk geometry being homogeneous and isotropic. 

It may appear puzzling how a homogeneous and isotropic bulk geometry can give rise to
a nonzero angular momentum, as directly applying the standard AdS/CFT dictionary 
to such a geometry will clearly yield a zero value. 
The key idea,  following~\cite{Liu:2012zm}, is to consider a small and slightly {\it inhomogeneous} perturbation $\delta \vartheta$ around the background value of
$\vartheta$, which results in a nonzero momentum current density $\de T_{0i}$.\footnote{We use  latin letters in the middle of the alphabet $(i,\ j,\ k,\ \ldots)$ to denote two-dimensional spatial indices on the boundary.}
To leading order in the derivative expansion along the boundary directions, $T_{0i}$ depends linearly on $\ep_{ij} \p_j \de \vt$. Now let us consider a configuration of $\de \vt $ which is homogeneous along boundary spatial directions inside a big box but vanishes outside. 
Then $\de T_{0i}$ is only nonvanishing at the edge of the box, but remarkably such an edge current generates an angular momentum proportional to the volume of the box  
\be 
\de J =  \ep_{ij} \int d^2 x \,x_i \, \de T_{0j} \propto V_{\rm box} \delta \vt
\ee
resulting in a nonzero angular momentum density $\de \sL$ which survives even when we take the size of the box to infinity. Thus in the homogeneous limit, the angular momentum density $\de \sL$ arises from the global effect of an edge current, which explains why it is not visible from the standard local analysis of the stress tensor.

When $\vt$ is dual to a marginal operator,  $\de \vt$ is independent of radial direction of AdS
and $\de \sL$ is given by $\de \vartheta$ times a constant which can be easily integrated to find 
the value of $\sL$ for a finite $\vt$. But for $\vt$ dual to a boundary relevant operator, $\de \vt$ has a nontrivial radial evolution (which simply reflects that a relevant operator flows), and the relation between $\de \sL$ and $\de \vt$ involves a somewhat complicated radial integral over various bulk fields. Remarkably, this relation can be written as a total variation in the space of gravity solutions, which can then be easily  
integrated to yield a closed expression for $\sL$ at a finite $\vt$.

 More explicitly, we consider a most general bulk metric consistent with translational and rotational 
 symmetries along boundary directions, which can be written in a form 
 \be \label{bmetric}
 ds^2 =  \frac{\ell^2}{z^2} \lb -f(z) dt^2 +h(z)dz^2 + \lb dx^i \rb^2 \rb \ 
\ee
with $z=0$ as the boundary. Matter fields include $\vt (z)$, $A_t (z)$, and possibly others. 
We denote $z_0$ as the horizon of the metric.  Note that in the coordinate choice of~\eqref{bmetric}
$z_0$ is inversely proportional to the square root of the entropy density $s$, i.e. $z_0 \propto s^{-\ha}$,
and serves as an IR cutoff scale\footnote{Physically, it can be interpreted as characterizing the correlation length of the boundary system.} of the boundary system.  
For the axionic coupling~\eqref{axio} we find that the angular momentum density can be written as  
\be \label{l1}
\sL = - \frac{2 \bCS \ell^2}{\kappa^2}\mu^2\vartheta(z_0)
+ \frac{2 \bCS \ell^2}{\kappa^2} \intop^{z_0}_0 dz \lb A_t(z) - \mu \rb^2  \vartheta'(z)
\ee
where $\mu$ is the chemical potential, $\ell$ is the AdS radius, and $\kappa^2 = 8 \pi G_4$.  
For gravitational CS coupling~\eqref{gcs}, we find that 
\be \label{l2}
\sL = - \frac{4\pi^2 \aCS \ell^2}{\kappa^2} T^2 \vartheta (z_0) + 
\frac{\aCS \ell^2}{4\kappa^2} \intop_0^{z_0} dz \lb\frac{f'^2}{fh}\rb \vartheta'
\ee
where $T$ is the temperature. 

Equations~\eqref{l1}--\eqref{l2} are universal in the bulk sense that 
they have the same form in terms of bulk gauge fields or metric components, independent of the specific form of bulk actions, geometries and possible other matter fields. But they are not universal in the boundary sense 
as it appears that they cannot be further reduced to expressions in terms of boundary quantities only.

When $\vartheta$ is dual to a marginal operator at the boundary, $\vt (z)$ is constant in the bulk and its value can be identified as the coupling of $\sO$. 
Then for both~\eqref{l1} and~\eqref{l2}, $\sL$ is given by the first term, reproducing our earlier results in  \cite{Liu:2012zm}. These expressions are now universal also in the boundary sense, valid for any boundary theory with a gravity dual.  
In Sec.~\ref{sectionIV}, we will present a preliminary explanation of this universal behavior from the perspective of the boundary conformal field theory (CFT).
We hope to explore this point in future.

For $\vt$ dual to a relevant operator, $\vt (z)$ can be interpreted as the running coupling for the corresponding boundary operator $\sO$, with $z$ as the renormalization group (RG) length scale. In this case, the first term of~\eqref{l1} and~\eqref{l2}  is proportional to the running coupling evaluated at the IR cutoff scale $z_0$. The second term 
of~\eqref{l1} and~\eqref{l2} has the form of the beta function (given by $\vt'$) for $\sO$ integrated over the RG trajectory all the way to the IR cutoff. This indicates that in the case of a relevant operator, despite being an IR quantity, the angular momentum receives contribution from all scales. The simplicity of the integration kernel in these equations may suggest a possible simple boundary interpretation which should be explored further.  

Another interesting phenomenon associated to parity violation in 
$2+1$ dimensions is the Hall viscosity \cite{Hallviscosone}. It turns out that,
in quantum Hall states, there is a close relation between the Hall viscosity 
and the angular momentum density \cite{Nicolis:2011ey,Hallviscostwo, Hallviscostwo1, Hallviscostwo2}.  
It would be interesting to understand how universal such a relation is. 
In a forthcoming paper, we will discuss this issue from the holographic 
perspective. We will apply the prescription of \cite{SaremiSon} 
to identify models where the Hall viscosity is nonzero and compare its
value with the angular momentum density.  

For the remainder of this paper, we will use the following. Latin letters stand for $(3+1)$-dimensional spacetime indices, greek letters stand for $(2+1)$-dimensional indices on the boundary, latin letters in the middle of the alphabet $(i,j,k,\ldots)$ stand for 2-dimensional spatial indices on the boundary and $\pd^2 \equiv \pd_x^2 + \pd_y^2$. The metric is denoted via $g_{ab}$ with signature $(-,+,+,+)$ in the bulk, and via $h_{\alpha\beta}$ on the boundary; the Einstein summation convention and geometric units with $\hbar = c = 1$ are assumed, unless otherwise specified; we denote $\kappa^2=8\pi G_4$.

After posting this paper on the arXiv e-print server, 
it was pointed out by K.~Landsteiner and by a referee of this paper that the spontaneous generation
of the edge current and of the angular momentum in $2+1$ dimensions discussed in this paper
may be related to the chiral magnetic effect \cite{chiralone,SonSurowka,landsteiner} and axial magnetic effect \cite{axialone,axialtwo,axialthree} in $3+1$ dimensions. Prompted by their suggestions, we found that the effects in $3+1$ dimensions 
and $2+1$ dimensions are indeed related by dimensional reduction when the parity-violating perturbation 
is {\it marginal}, which was the focus of our previous paper \cite{Liu:2012zm}. 
For completeness, we added Sec.~\ref{sectionIV} to discuss the relation.  
The purpose of this paper is to generalize our results to the case when the parity-violating perturbation is {\it relevant}, and
the discussion in Sec.~\ref{sectionIV} is not immediately applicable.
There may exist a generalization of the chiral magnetic effect and axial magnetic effect in 
$3+1$ dimensions which correspond to dimensional oxidation of the effects studied in this paper.

\section{Axionic coupling}
\label{secIIAx}


In this section we consider a scalar field $\vartheta$ coupled to a Maxwell field via an axionic coupling, $\vartheta \SF^{ab} F_{ab}$. We first explicitly work out the angular momentum for a simple setup, and then generalize 
the results to general gravity theories.

\subsection{\label{secangmom} Angular momentum}

\subsubsection{Small perturbations}

Consider the action
\ba
S &=& \frac{1}{2\kappa^2}\intop d^4x \sqrt{-g} \bigg[ R - \frac{1}{2}\lb \pd \vartheta \rb^2 - V(\vartheta) \nn\\
&-& \ell^2 F^{ab}F_{ab} - \ell^2 \bCS \vartheta \SF^{ab} F_{ab} \bigg], 
\label{axac}
\ea
with $\bCS$ a coupling constant, and $\vt$ dual to a relevant (or marginal) pseudoscalar boundary operator.  We assume that the background geometry
is asymptotically AdS with $\ell$ the AdS radius. 
The equations of motion are
\ba
& & R_{ab} - \frac{1}{2}\pd_a\vartheta\pd_b\vartheta -\frac{1}{2}g_{ab}V(\vartheta) \nn
\\& & 
- 2\ell^2\lb F_{ca}F^c_{\ b} - \frac{1}{4}g_{ab}F^2 \rb = 0,\\
& & \frac{1}{\sqrt{-g}}\pd_a\lb g^{ab}\sqrt{-g} \pd_b \vartheta \rb - V'(\vartheta) 
- \bCS \ell^2 \SF F = 0 \nn \\
\\
& & \pd_a\lsb \sqrt{-g} \lb F^{ab} + \bCS \vartheta \SF^{ab} \rb \rsb = 0.
\ea
A most general solution describing the boundary in a static, homogeneous, isotropic 
state 
can be written as 
\ba
\label{metric00}
& &g^{(0)}_{ab} dx^a dx^b = \frac{\ell^2}{z^2} \lb -f(z) dt^2 +h(z)dz^2 + \lb dx^i \rb^2 \rb, \cr
& &\vartheta = \vartheta(z),\qquad A_a = A_t(z)\delta_a^t.
\ea
The AdS boundary lies at $z=0$ 
with 
\be \label{asy1}
f(z)  \to  1 , \qquad h(z) \to 1 , \quad z \to 0
\ee
and 
\be 
A_t (z=0) = \mu
\ee
where $\mu$ is the chemical potential. We assume that there is a horizon as $z=z_0$, where 
$f (z)$ has a simple zero and $h(z)$ has a simple pole. The temperature 
is given by 
\be
T = {1 \ov 4 \pi} \sqrt{f'(z_0) h^{-1'} (z_0)}  .
\ee

Here are some background equations of motion which will be important below.  
The $t$ component of the background Maxwell equation can be integrated to give,
\be
\label{Atis}
A_t'(z) = Q \sqrt{f(z)h(z)}
\ee
with $Q$ the charge density. 
The $tt$ and $ii$ components of the background Einstein equations can be used to obtain
\be
\label{KAT0}
4 \sqrt{fh} Q^2 = \le({f' \ov z^2 \sqrt{fh}}\ri)' .
\ee



As discussed in the Introduction, to compute the angular momentum, we consider a small and slightly inhomogeneous perturbation $\de \vt (z, x^i)$
around the background value $\vt (z)$. Such a perturbation will clearly also induce perturbations of 
the metric and gauge field, 
\ba
\label{metric}
g_{ab} &=& g^{(0)}_{ab} +   \frac{\ell^2}{z^2} \delta g_{ab},\\
\label{bckgAa}
A_a &=& A_t(z)\delta_a^t + \delta\! A_a(z,x^i) .
\ea
The metric and gauge field perturbations will be assumed to be normalizable, while $\de \vt$ can 
be either normalizable or non-normalizable. We will also make the following gauge choice,
\be
\label{gaugechoice29}
\delta\! A_z=0, \qquad \delta g_{zt}=0.
\ee

To find the angular momentum, we first compute $T_{ti}$, which in turn requires us to find $\de g_{ti}$. 
Since $\de \vt$ is small we can work at the linear order in all perturbations, 
and since we will eventually take $\de \vt$ to be homogeneous, it will be enough to keep only terms with at most one boundary spatial derivative (for details on the derivative expansion in holographic fluid dynamics see for instance \cite{hydrotwo,hydrofour}).  


We now proceed with the computation in detail. The $ti$ component of the Einstein equations reads
\be
\label{llhs13}
\left(\frac{f'}{f}+ \frac{h'}{h} +\frac{4}{z}\right)
   \pd_z \delta g_{ti}-2 \pd^2_z \delta g_{ti}
= 8 z^2 A_t' \pd_z \de A_i 
\ee
while the $i$ component of the Maxwell equations reads
\ba
\label{maxi3}
& & 
\pd_z\lb\frac{\sqrt{f}  \pd_z\delta\!A_i}{\sqrt{h}} +Q \delta g_{ti} \rb \cr
&+&
\bCS \epsilon_{ij} \lb \sqrt{fh} Q \pd_j\delta\vartheta - \vartheta' \pd_j \delta\!A_t \rb  = 0 .
\ea
Due to the presence of $\de \vt$ and $\de A_t$ in~\eqref{maxi3}, 
Eqs.~\eqref{llhs13}--\eqref{maxi3} do not close between themselves, which implies that 
solving $\de g_{ti}$ explicitly will be a very complicated task, if possible at all.\footnote{Note that equations 
for $\de \vt$ and $\de A_t$ are rather complicated. This is especially the case for more general action~\eqref{newacti}.} Fortunately as we will see it turns out to be unnecessary to do so.


Integrating \eqref{maxi3} from the horizon to $z$ we find that
\ba
&& \pd_z\delta\!A_i(z,x^k) = \frac{\sqrt{h(z)}}{\sqrt{f(z)}}\Bigg[ - Q\delta g_{ti}(z,x^k)
\nn \\ &&  + \bCS \epsilon_{ij} \intop_{z0}^z dw \, \le(\vartheta' \pd_j \delta\!A_t 
- A_t' \pd_j \delta\vartheta  \ri) \Bigg] 
\label{Awewant}
\ea
where we have assumed that $\pd_z \delta\!A_i(z,x^k)$ is nonsingular at the horizon.

Plugging Eqs.~\eqref{Awewant} into~\eqref{llhs13}, and using~\eqref{KAT0} we find that
\ba
& &\pd_z \lsb\frac{f^\frac{3}{2}(z)}{z^2\sqrt{h(z)}}\pd_z\lb\frac{\delta g_{ti}(z,x^k)}{f(z)}\rb\rsb  \\
&=& 4\bCS \epsilon_{ij} A_t'(z) \intop_{z_0}^z dw \lsb A_t' \pd_j \delta \vartheta - \vartheta' \pd_j \delta A_t \rsb. \nn
\label{eq25}
\ea
The above equation implies that despite the mixing between $\de A_i$ and $\de g_{ti}$, 
the combination ${1 \ov f} \de g_{ti}$ remains ``massless.''  Writing $g_{ab} = g^{(0)}_{ab} + g^{(1)}_{ab}$
with $ g^{(1)}_{ab} = {\ell^2 \ov z^2} \de g_{ab}$, we note that ${1 \ov f} \de g_{ti}$ in fact corresponds to $(g^{(1)})^t_i$.


Integrating Eq.~\eqref{eq25} from the boundary $z=0$ to the horizon $z_0$, we find that 
\ba
& &\frac{f^\frac{3}{2}(z)}{z^2\sqrt{h(z)}} \pd_z\lb \frac{\delta g_{ti}(z,x^k)}{f(z)} \rb{\Bigg |}_{z=0} = 4\bCS \epsilon_{ij} \times \nn\\
& &\times \intop_{z_0}^0 dz A_t'(z) \intop_{z_0}^z dw \lsb A_t' \pd_j \delta \vartheta - \vartheta' \pd_j \delta A_t \rsb
\label{nne}
\ea
where we have used that at the horizon
\be \label{hoe}
\de g_{ti} (z_0, x^i) = 0
\ee
and $\p_z \de g_{ti}$ is regular there. Equation~\eqref{hoe} is analogous to the well-known statement that $A_t$ vanishes at black hole horizons, and is similarly most transparent in Euclidean signature, where a nonzero $\de g_{ti}$ at the shrinking time cycle indicates a delta-function contribution to the Einstein tensor. It can be also shown directly from consistency of various components of Einstein equations (see Appendix~\ref{app:horb}). 

Now consider the left-hand side of~\eqref{nne}. With $\delta g_{ti}$ normalizable, i.e. 
\be 
\label{asy4}
\delta g_{ti}(z,x^l) = G^{(3)}_i(x^l)z^3 + \mathcal{O}(z^4).
\ee
we find 
\ba
& &3G^{(3)}_i = 4\bCS \epsilon_{ij} \intop_{z_0}^0 dz A_t'(z) \intop_{z_0}^z dw  \times \\
& &\times \lsb A_t'(w) \pd_j \delta \vartheta(w) - \vartheta'(w) \pd_j \delta A_t(w) \rsb.\nn
\ea
Using the standard formulas as in~\cite{KB1999, BFS2002, dHSS2001} (see also Appendix~\ref{sec:app_b} and the Appendix of~\cite{Liu:2012zm}), the boundary stress-energy tensor is
\be
\label{Tti}
\de T_{ti} = \frac{3\ell^2}{2\kappa^2}G^{(3)}_i  = - \ep_{ij} \p_j \de \Phi
\ee
where 
\ba 
\de \Phi &=& \frac{2\bCS\ell^2}{\kappa^2}\intop^{z_0}_0 dz A_t'(z) 
\intop_{z_0}^z dw \lsb A_t' \delta \vartheta - \vartheta' \delta A_t \rsb \cr
& = & \frac{2\bCS\ell^2}{\kappa^2}
\intop_{z_0}^0 dw \lsb A_t' \delta \vartheta - \vartheta' \delta A_t \rsb (A_t - \mu) .
\label{deph}
\ea
In the second equality above we have exchanged the order of integration to perform one integral and used
that $A_t(0)=\mu$.

Now consider a configuration of $\de \vt $ which is homogeneous along boundary spatial directions inside a big box but vanishes outside. The above $\de T_{0i}$ is nonvanishing only at the edge of the box, but generates an angular momentum proportional to the volume of the box, resulting in an angular momentum density 
\be 
\delta\mathcal{L}   = 2 \de \Phi 
\label{gne}
\ee
We now take the box size to infinity, with $\de \vt$ and $\de A_t$ homogeneous everywhere with no dependence 
on $x^i$.



\subsubsection{Angular momentum density}

Equations~\eqref{Tti}--\eqref{gne} apply to infinitesimal variations $\de \vt$ and $\de A_t$  around~\eqref{metric00}. To compute $\sL$ for~\eqref{metric00}, we need to integrate~\eqref{deph} 
along some trajectory 
in the space of field configurations from a configuration with $\vt =0$ (and thus $\sL =0$) 
to~\eqref{metric00}, i.e. schematically 
\be
 \Phi =  \int_{\vt =0}^{\vt} \de \Phi 
 \ee
from which we then find  
\be \label{gdj}
T_{ti} = - \ep_{ij} \p_j \Phi, \qquad \sL = 2 \Phi .
\ee   
At first sight this appears to be an impossible task as solving $\de A_t$ in terms of $\de \vt$ is complicated 
and so is integration over field space as $A_t$ in general also has nontrivial $\vt$ dependence. 

Remarkably, Eq.~\eqref{deph} can be written as a total derivative $\delta$ in the field configuration space.
Choosing a trajectory in configuration space with a fixed $\mu$ (i.e. $\de \mu =0$)
we can rewrite~\eqref{deph}  as 
\ba
\de \Phi &=&  \frac{\bCS\ell^2}{\kappa^2}
 \intop_{z_0}^0 dw \lsb B' \delta \vartheta - \vartheta' \delta B \rsb \cr
& = &  \frac{ \bCS\ell^2}{\kappa^2} \intop_{z_0}^0 dw \le[ (B \de \vt)' - \de (B  \vt')\ri]
\label{enn}
\ea
where $B = A_t^2 - 2 \mu A_t$ and 
 in the second line we have used that for arbitrary functions $F$ and $G$
\be
\label{eqaux}
\lb F\delta G\rb' - \delta \lb F G' \rb = F'\delta G - \delta F G' .
\ee
Recall that $A_t$ is zero at the horizon and equal to $\mu$ at the boundary. 
Evaluating the total derivative and taking $\de$ operation outside the integral for the second term, 
Eq.~\eqref{enn} becomes 
\be \label{bne}
\de \Phi = \frac{ \bCS\ell^2}{\kappa^2} \de  \le[- \mu^2  \vt (0) + \int_0^{z_0} 
dw \, (A_t^2 - 2 \mu A_t) \vt' \ri] .
\ee
Note that in exchanging the order of $\de$ with the integration, there is a term proportional to $\de z_0$, which, however, vanishes as $A_t(z_0)=0$. Now~\eqref{bne} is a total variation and we conclude that 
\be \label{bne1}
\Phi =  \frac{ \bCS\ell^2}{\kappa^2} \le[- \mu^2  \vt (0) + \int_0^{z_0} 
dw \, (A_t^2 - 2 \mu A_t) \vt' \ri]  .
\ee
The above equation can also be slightly rewritten as 
\be
 \Phi = \frac{\bCS\ell^2}{\kappa^2} \le[- \mu^2  \vt (z_0) + \int_0^{z_0} 
dw \, (A_t - \mu)^2 \vt' \ri] .
\label{bne2}
\ee
Note that Eqs.~\eqref{bne1}--\eqref{bne2} also apply to inhomogeneous configurations 
as far as the spatial variations are sufficiently small.

When $\vt$ is dual to a marginal operator,  $\vartheta$ is constant in the bulk with $\vt (0) = \vt (z_0) = \vt$,
and the second term in~\eqref{bne1} or~\eqref{bne2} drops out. We then recover the result of~\cite{Liu:2012zm}, 
\be
\label{marginalcase}
 {\cal L} = - \frac{2\bCS\ell^2}{\kappa^2}\mu^2\vartheta .
\ee
For a general relevant operator, the second term in~\eqref{bne1} or~\eqref{bne2} is nonzero and the angular momentum density will receive contribution from integration over the bulk full spacetime. In terms of boundary 
language, the angular momentum receives contributions from degrees of freedom at all scales. 
Also note that for a relevant operator $\vt (0) =0$, so in~\eqref{bne1} the sole contribution comes from
the second term.

\subsubsection{An explicit example}
\label{subsec:ex}

We now consider an explicit example. For simplicity we take $V (\vt) = \ha m^2 \vt^2$ 
with $m^2=-2$. Thus $\vt$ is dual to a relevant boundary operator $\sO$ in $d=3$ with $\De = 2$. 
We will consider a solution~\eqref{metric00} in which $\vt$ is non-normalizable, i.e. $\vt$ has the asymptotic 
behavior near the boundary 
\be 
\vt (z) = M z + O(z^2), \qquad z \to 0
\ee
where $M$ is a parameter of dimension mass. The solution~\eqref{metric00} then describes a boundary theory flow upon turning on a relevant perturbation $\int d^3 x \, M \sO$, with $M$ interpreted as the bare coupling. 
Since we are considering the system at a finite density/finite temperature, the flow is cut off at some infrared scale 
characteristic of finite density/finite temperature physics. In the coordinate system we are using in~\eqref{metric00}, such a scale should correspond to location of the horizon $z_0 \propto s^{-\ha}$ with $s$ the entropy density.

\begin{figure}
\centering
\includegraphics[width=8.5cm,clip=true]{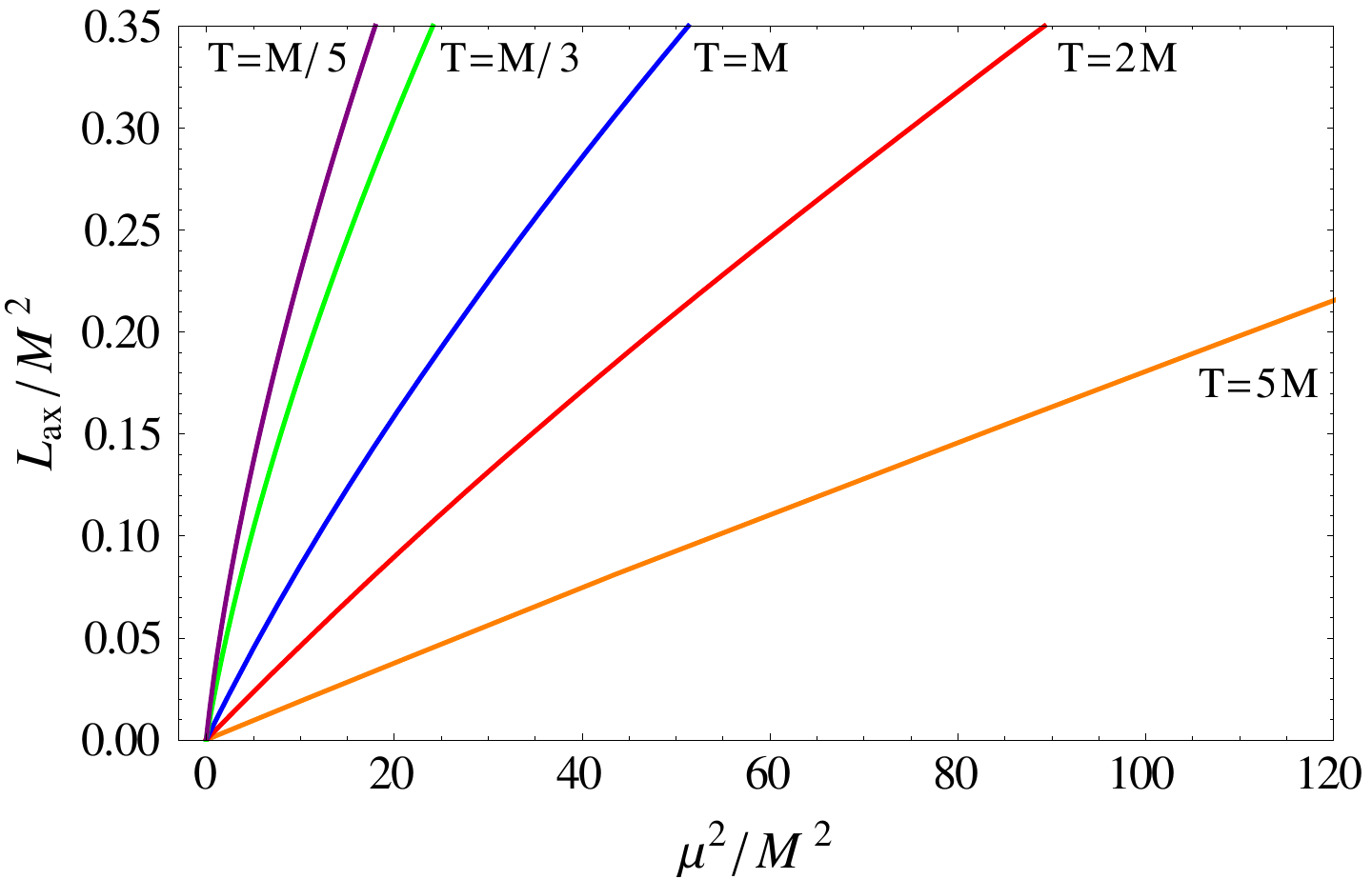}
\caption{(Color online) Angular momentum density as a function of $\mu^2/M^2$ for axionic coupling and non-normalizable scalar field in a quadratic potential with $m^2=-2$.}
\label{fig:AX_L}
\end{figure}

\begin{figure}
\centering
\includegraphics[width=8.5cm,clip=true]{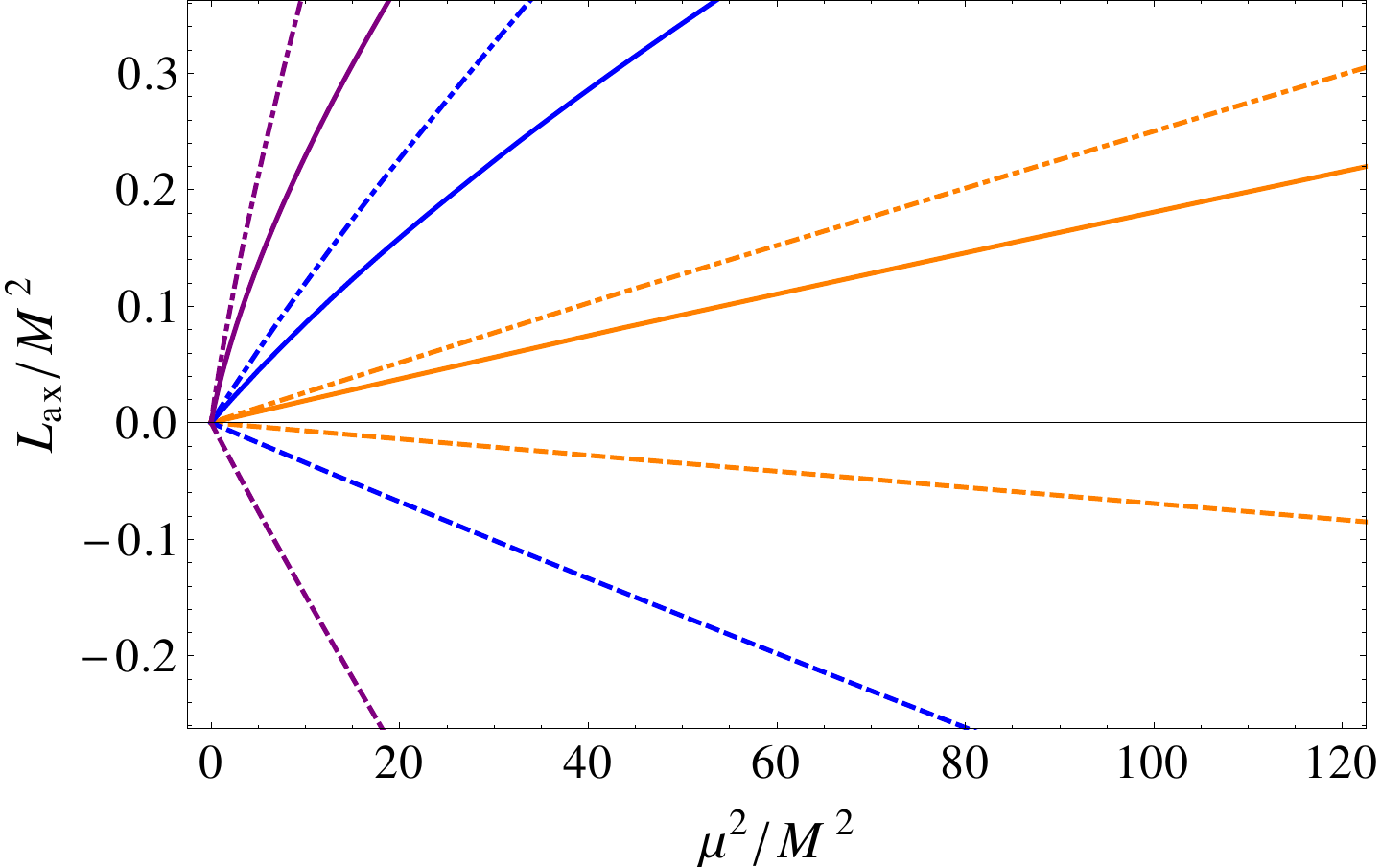}
\caption{(Color online) Angular momentum density as a function of $\mu^2/M^2$ for axionic coupling and non-normalizable scalar field in a quadratic potential with $m^2=-2$ for $T=5M$ (orange), $T=M$ (blue) and $T=M/5$ (purple). The total angular momentum is represented by solid lines, the first term in Eq. \eqref{l1} by dot-dashed lines and the second term in Eq. \eqref{l1} by dashed lines.}
\label{fig:AX_L_OVER}
\end{figure}

\begin{figure}
\centering
\includegraphics[width=8.5cm,clip=true]{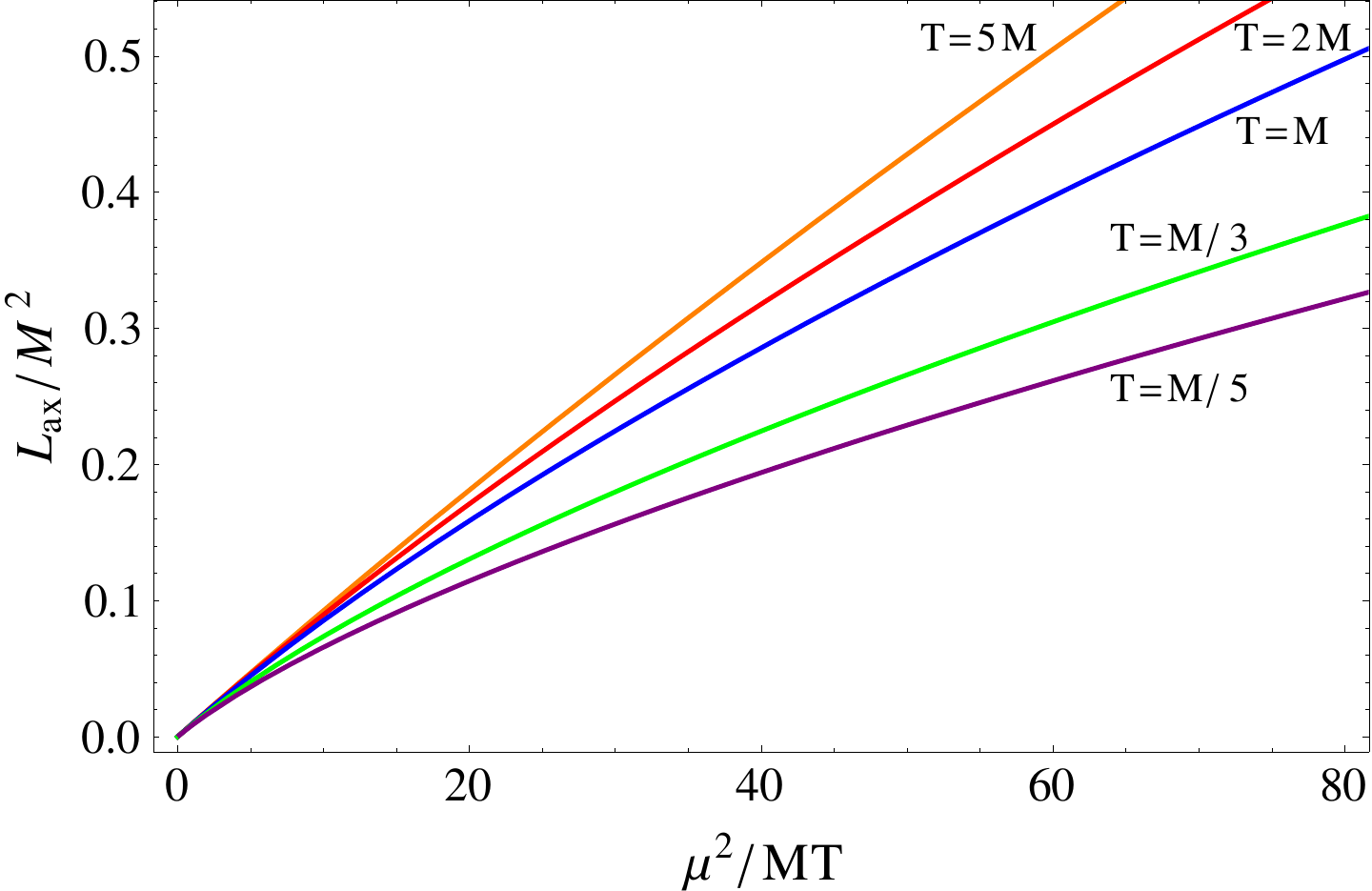}
\caption{(Color online) Angular momentum density as a function of $\mu^2/MT$ for axionic coupling and non-normalizable scalar field in a quadratic potential with $m^2=-2$.}
\label{fig:AX_L_ALT}
\end{figure}

\begin{figure}
\centering
\includegraphics[width=8.5cm,clip=true]{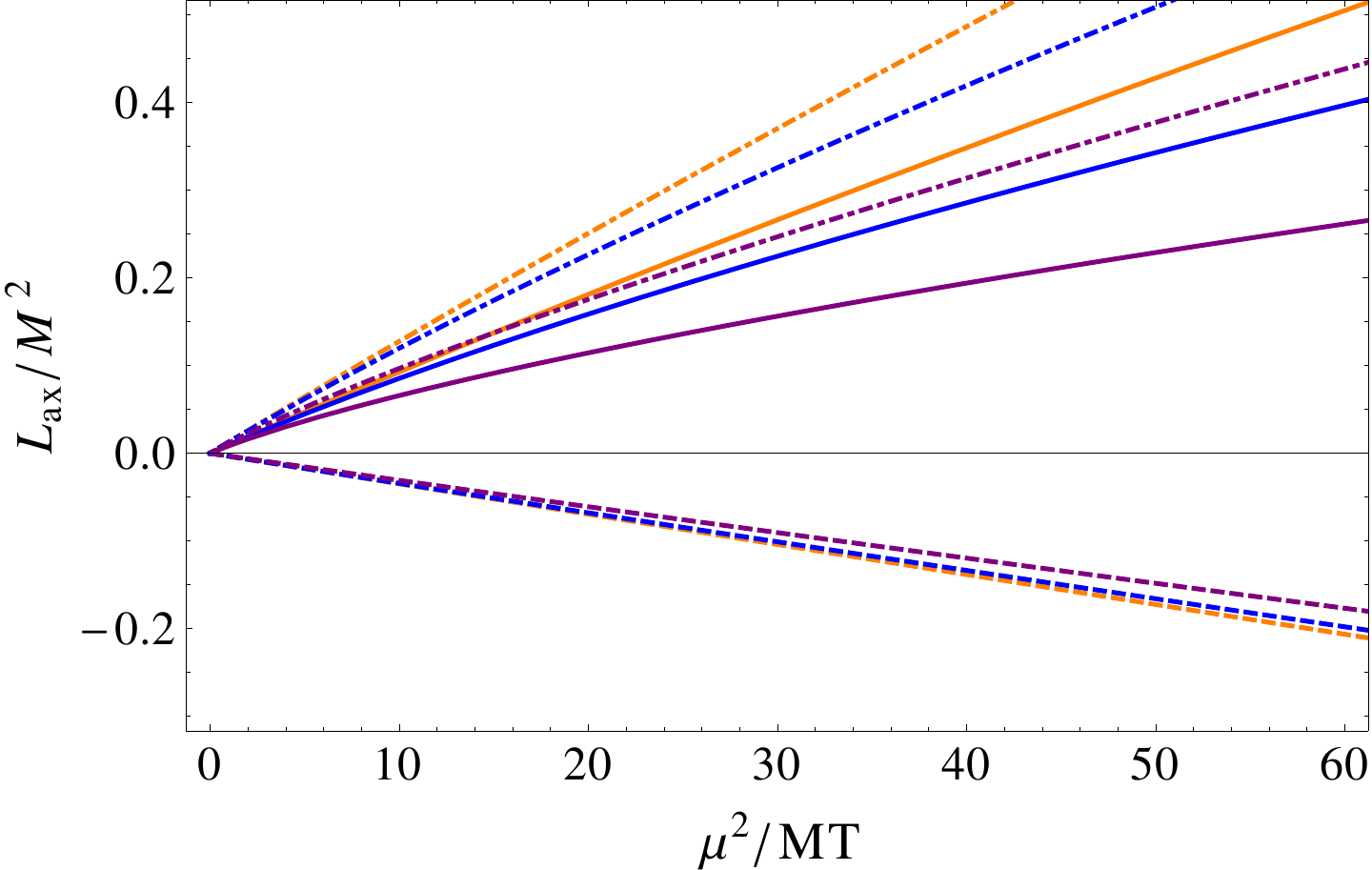}
\caption{(Color online) Angular momentum density as a function of $\mu^2/MT$ for axionic coupling and non-normalizable scalar field in a quadratic potential with $m^2=-2$ for $T=5M$ (orange), $T=M$ (blue) and $T=M/5$ (purple). The total angular momentum is represented by solid lines, the first term in Eq. \eqref{l1} by dot-dashed lines and the second term in Eq. \eqref{l1} by dashed lines.}
\label{fig:AX_L_OVER_ALT}
\end{figure}

We present plots of the axionic angular momentum as a function of $\mu^2/M^2$ in Figs. \ref{fig:AX_L} and \ref{fig:AX_L_OVER} and as a function of $\mu^2/MT$ in Figs. \ref{fig:AX_L_ALT} and \ref{fig:AX_L_OVER_ALT}. We exhibit the two terms entering Eq. \eqref{l1}, as well as the total angular momentum, in Figs. \ref{fig:AX_L_OVER} and \ref{fig:AX_L_OVER_ALT}. We note that in the large $T$ regime the angular momentum density grows as $L_\mrm{ax} \propto \mu^2M/T$. This is expected from the general structure of Eq. \eqref{l1} since roughly speaking the angular momentum is proportional to $A_t^2$ and $\vartheta$, while the gauge field is proportional to $\mu^2$ plus corrections and the scalar field is proportional to $M/T$ plus corrections. When $T\rightarrow 0$, the angular momentum tends to a finite constant. We also remark that out of the three contributions represented in Figs. \ref{fig:AX_L_OVER} and \ref{fig:AX_L_OVER_ALT}, the second term in Eq. \eqref{l1} varies almost linearly with $\mu^2/MT$ over the interval we have considered.

\subsection{Electric edge current}

Another interesting phenomenon associated to the axionic coupling is a spontaneous generation of
the electric current dual to the bulk gauge field. As we will see in Sec. \ref{sectionIV}, this is closely related to the
angular momentum generation when the scalar field $\vartheta$ is dual to a marginal operator. 

The expectation value of the current is defined in terms of the normalizable mode of the bulk gauge field as
\be
\delta j_i = -\frac{4\ell^2}{2\kappa^2} \lim_{z\rightarrow0}  \frac{\delta A_i}{z} = -\frac{4\ell^2}{2\kappa^2} \lb\delta A_i \rb'\Big|_{z=0}.
\label{normalizablemode}
\ee
Evaluating \eqref{Awewant} at the boundary we obtain (with $\delta g_{ti}$ normalizable)
\be
\pd_z\delta\!A_i(z=0,x^k)
= - \bCS \epsilon_{ij} \pd_j \intop^{z_0}_0 dw \Big[  \vartheta'  \delta\!A_t - A_t' \delta\vartheta \Big] 
\ee
leading to 
\be 
\delta j_i =  - \ep_{ij} \p_j \de\chi
\ee
with 
\be 
\de \chi = \frac{2\ell^2 \bCS }{\kappa^2}  \intop^{0}_{z_0} dw \Big[  \vartheta'  \delta\!A_t - A_t' \delta\vartheta \Big] .
\ee
Again using~\eqref{eqaux}, the above equation can be written as a total variation in the space of configurations
\ba 
\delta \chi &=& - \frac{2\ell^2 \bCS }{\kappa^2} \delta\! \lsb  \mu \vt  (0)+ 
\intop^{z_0}_0 dw \, \vartheta' A_t \rsb \cr
&=&   \frac{2\ell^2 \bCS }{\kappa^2}  \delta\! \lsb  
\intop^{z_0}_0 dw \, \vartheta A_t' \rsb .
\ea
We thus find an electric  current 
\be 
j_i = - \ep_{ij} \p_j \chi
\label{edgeone}
\ee
with
\be 
\chi = \frac{2\ell^2 \bCS }{\kappa^2}  \lsb  
\intop^{z_0}_0 dw \, \vartheta A_t' \rsb .
\label{edgetwo}
\ee


\subsection{Bulk universality}
\label{secunivers}

The results obtained in the previous subsections extend without modification to most general two-derivative 
theories of the form 
\ba
S &=& \frac{1}{2\kappa^2}\intop d^4x \sqrt{-g} \bigg[ R  
- \frac{1}{2} G^{IJ}(\vartheta^K)\pd_a \vartheta^I \pd^a \vartheta^J  \nn\\
&-& V(\vartheta^K) - \ell^2 Z^{PQ}(\vartheta^K) F^P_{ab}F^{Qab} \nn\\
&-& \ell^2 \bCS C^{PQ}(\vartheta^K) \SF^{Pab} F^Q_{ab} \bigg] .
\label{newacti}
\ea

In the above $I,\ J,\ K$ label different scalar fields, while $P, Q$ label different vector fields, and $G^{IJ}$, $Z^{PQ}$ and $C^{PQ}$ are functions of scalar fields $\vt^K$. They are symmetric and assumed to be invertible. 
We consider a metric of the form~\eqref{metric00} with 
\be 
\vartheta^I = \vartheta^I(z), \quad
 A^P_a =  A^P_t(z) \de_a^t, \quad     A^P_t(0) = \mu^P  
\ee
where $\mu^P$ is the chemical potential for boundary conserved current $J^P$ 
dual to $A^P_a$. 

The discussion exactly parallels that of Sec.~\ref{secangmom} so below we will simply list 
the counterparts of the key equations there.

Background equations of motion~\eqref{Atis}--\eqref{KAT0} now become 
\be
\label{Atis1}
{A_t^P}'(z) = (Z^{-1})^{PR}Q^R \sqrt{f(z)h(z)}
\ee
with $Q^R$ the charge density for $J^R$ and 
\be
\label{KAT1}
4 \sqrt{fh} (Z^{-1})^{PR}Q^PQ^R = \le({f' \ov z^2 \sqrt{fh}}\ri)' .
\ee

As before we consider general small perturbations generated by a small and slow-varying $\de \vt^I (z,x^i)$ and make the gauge choice
\be
 \delta\! A^P_z=0, \qquad \delta g_{zt}=0 .
\ee
Equations~\eqref{llhs13} and~\eqref{maxi3} then generalize respectively 
to
\be
\label{llhs14}
\left(\frac{f'}{f}+ \frac{h'}{h} +\frac{4}{z}\right)
   \pd_z \delta g_{ti}-2 \pd^2_z \delta g_{ti}
= 8 z^2 Z^{PQ} {A_t^P}'
   \pd_z A^Q_i
\ee
and 
\ba
\label{maxi4}
& & 
\pd_z\lb\frac{\sqrt{f(z)}Z^{PQ}\pd_z \delta A^Q_i}{\sqrt{h(z)}} +Q^P \delta g_{ti} \rb \\
&+&
\bCS \epsilon_{ij} \pd_j  \lb \delta C^{PQ} {A_t^Q}'  -{C^{PQ}}' \delta A_t^Q   \rb  = 0 .\nn
\ea

Then identical manipulations as before lead to~\eqref{gdj} with 
\ba 
&& \Phi = \frac{\bCS\ell^2}{\kappa^2} \biggl[- \mu^P\mu^Q C^{PQ}(0)  \biggr. \\
&& \biggl. + \intop^{z_0}_0 dw \lb  A_t^PA_t^Q(w) - 2 \mu^P A^Q_t(w)\rb{C^{PQ}}'(w)  \biggr] \nn
\ea
or equivalently 
\ba 
&& \Phi = \frac{\bCS\ell^2}{\kappa^2} \biggl[- \mu^P\mu^Q C^{PQ}(z_0)  \biggr. \\
&& \biggl. + \intop^{z_0}_0 dw \le(A_t^P - \mu^P \ri) \le(A_t^Q - \mu^Q \ri) {C^{PQ}}'  \biggr] .\nn
\ea

\section{Gravitational Chern-Simons Term}

In this section, we consider the induced stress tensor and angular momentum density for bulk theories 
where parity violation is generated by the gravitational Chern-Simons coupling, $\vartheta \SR R$.  We will first consider a simple example with a relevant scalar operator and then generalize the discussion 
to generic theories. The discussion is similar to that of the last section, so we will be briefer.


\subsection{Relevant scalar field}
\noindent Consider the action
\be \label{gcsac}
S = \frac{1}{2\kappa^2}\intop d^4x \sqrt{-g} \bigg[ R - \frac{1}{2}\lb \pd \vartheta \rb^2 - V(\vartheta)
- \frac{\aCS \ell^2}{4}\vartheta \SR R \bigg]
\ee
where $\aCS$ is a constant and $\vt$ is dual to a relevant (or marginal) pseudoscalar boundary operator. In~\eqref{gcsac}
\be
 \SR R = \SR^{abcd} R_{bacd}, \quad \SR^{abcd} = \ha \ep^{cdef} R^{ab}{_{ef}} 
 \ee
and  $\ep^{abcd}$ is the totally antisymmetric tensor with  
 $\epsilon^{012z}={1/\sqrt{-g}}$.
The equations of motion are
\ba
& & R_{ab} - \frac{1}{2}\pd_a\vartheta\pd_b\vartheta -\frac{1}{2}g_{ab}V(\vartheta) 
= \aCS \ell^2 C_{ab},\nn\\
& & \frac{1}{\sqrt{-g}}\pd_a\lb g^{ab}\sqrt{-g} \pd_b \vartheta \rb -\frac{\aCS\ell^2}{4}\SR R = 0 \  
\ea
where $C^{ab} \equiv \nab_c (\nab_d \vt \SR^{c(ab)d})$. 

We again consider a  solution 
of the form~\eqref{metric00} (without the gauge field). 
The strategy is the same as before. 
We consider a small and slowly varying perturbation $\de \vt (z, x^i)$ and work out the momentum 
response $\de T_{ti}$ to order $O(\ep)$ where the power $\ep$ counts the number of spatial 
derivatives of $\de \vt$. 
 We then write the resulting expression 
as a total variation in the space of field configurations which enables us to find the angular momentum 
associated with~\eqref{metric00}.

We will choose a gauge where $\de g_{tz} = \de g_{xx} = \de g_{yy} =0$. 
With the Einstein equations schematically reading
\be
\mrm{LHS}_{ab} = \aCS \ell^2 C_{ab}
\ee 
we note that in this gauge,
\be
\label{LHSnowis}
\mrm{LHS}_{ti} = \frac{-z^2}{2\sqrt{f(z)h(z)}} \pd_z \lsb\frac{f^\frac{3}{2}(z)}{z^2\sqrt{h(z)}}\pd_z\lb\frac{\delta g_{ti}(z,x^k)}{f(z)}\rb\rsb 
\ee
and $\de g_{zi}$, $\de g_{xy}$ are all at least of order $O(\ep)$. We then find to $O(\ep)$, $C_{ti}$ can be written as 
\be \label{cti}
C_{ti} =  \frac{z^2}{\ell^2\sqrt{fh}}\epsilon_{ij}\pd_j \de\Psi
\ee
with 
\be 
\de \Psi = K' + \le( \frac{f'^2}{8 f h} \ri)' \delta \vartheta -\frac{f'^2\vartheta'\delta g_{tt}}{8f^2h}
+\frac{f'\vartheta'\delta g_{tt}'}{4fh} + \frac{f'^2\vartheta'\delta g_{zz}}{8fh^2} 
\ee
and 
\ba 
K &=& \frac{ff'h'+ h\left(f'^2- 2ff''\right)}{8 fh^2}\delta \vartheta
- \frac{f'\vartheta'\delta g_{zz}}{8 h^2} \nn \\
&+&\frac{f'\vartheta'\delta g_{tt}}{8fh} - \frac{\vartheta'\delta g_{tt}'}{4h} .
\label{kk}
\ea
Then following similar manipulations as in~\eqref{nne}--\eqref{Tti} we find that 
\be 
\de T_{ti} = - \ep_{ij} \p_j \de \Phi 
\ee
with 
\be 
\de \Phi =   {\aCS \ell^2 \ov \kappa^2} \intop_{z_0}^0 \de \Psi dz .
\ee
Note that $\de g_{tt} = - \de f$ and $\de g_{zz} = \de h$ and $\de \Psi$ can be further written as 
\be 
\de \Psi = K' + \lb\frac{f'^2\delta\vartheta}{8fh}\rb'  - \de \le(\frac{f'^2}{8 f h} \vartheta' \ri) .
\ee
It can then be immediately checked that the boundary terms coming from $K$ are all zero with the assumption of the asymptotic behavior
\be \label{ayep}
f(z) = 1 + \# z^{2 + 2 \al} + \cdots, \quad h(z) = 1 + \# z^{2 \beta} + \cdots
\ee
where $\alpha>0$, $\beta>0$. We then note further that 
\ba 
&& \intop^{z_0}_0 dz \, \le[ \lb\frac{f'^2\delta\vartheta}{8fh}\rb'  - \de \le(\frac{f'^2}{8 f h} \vartheta' \ri) \ri] \\
&& = \delta\lb \intop_0^{z_0} dz \lb\frac{f'^2}{8fh}\rb'\vartheta \rb 
- \delta \lb \frac{f'^2}{8fh}\bigg|_{z_0} \rb\vartheta(z_0) \nn
\ea
where the second term is proportional to  $\delta T$, and thus vanishes if we choose a path in configuration space such that $\delta T=0$. Collecting the above we thus find $\de \Phi$ is a total variation with 
\ba
\Phi& = &  - {\aCS \ell^2 \ov \kappa^2}  \intop_0^{z_0} dz \lb \frac{f'^2}{8fh}\rb'\vartheta  \\
& = & - \frac{2 \pi^2\aCS\ell^2}{\kappa^2} T^2 \vartheta (z_0) + 
\frac{\aCS\ell^2}{8\kappa^2} \intop_0^{z_0} dz \lb\frac{f'^2}{fh}\rb \vartheta'  \nonumber \\
\label{hne}
\ea
The angular momentum is thus given by 
\be 
\sL = 2 \Phi .
\ee
For a marginal $\vt$, $\vt$ is independent of $z$ and only the first term in~\eqref{hne} is present. We then find a universal result which is independent 
of specific forms of $f$ and $h$ 
\be
\sL = - \frac{4 \pi^2\aCS\ell^2}{\kappa^2} T^2 \vartheta .
\ee

\subsection{Generalizations}

The above discussion can be immediately generalized to theories of the form 
\ba
&& S= \frac{1}{2\kappa^2}\intop d^4x \sqrt{-g} \lsb R 
- \frac{1}{2} G^{IJ}(\vartheta^K)\pd_a \vartheta^I \pd^a \vartheta^J  \ri. \cr
&& \le. - V(\vartheta^K) - \ell^2 Z^{PQ}(\vartheta^K) F^P_{ab}F^{Qab} - {\aCS \ell^2 \ov 4} C (\vartheta^K) \SR R \rsb.
\nonumber \\ 
\ea
Fixing the gauge $A_z^P = 0$, one finds that 
\be
\label{Aiis}
\pd_z\delta\!A_i^P (z,x^k) = - Q^P \sqrt{\frac{h(z)}{f(z)}}\delta g_{ti}(z,x^k) .
\ee
From~\eqref{Aiis} one then finds that the $ti$ component of the Einstein equations can again be written as 
\be 
{\rm LHS}_{ti} =  \aCS \ell^2 C_{ti}
\ee
with ${\rm LHS}_{ti}$ given by~\eqref{LHSnowis} and $C_{ti}$ by~\eqref{cti}--\eqref{kk} except that 
everywhere in $C_{ti}$ the pseudoscalar $\vt$ is replaced by $C (\vt^I)$. In this case we thus find that
\ba
\Phi &=&  - {\aCS \ov \kappa^2}  \intop_0^{z_0} dz \lb \frac{f'^2}{8fh}\rb' C(\vt^I)  \\
&=& - \frac{2 \pi^2\aCS\ell^2}{\kappa^2} T^2 C (\vt^I (z_0)) \nn\\
&& + \frac{\aCS\ell^2}{8\kappa^2} \intop_0^{z_0} dz \lb\frac{f'^2}{fh}\rb C(\vartheta^I)'
\label{hne}
\ea

We also note in passing that in this case there is no electric edge current as 
\be
\delta j_i^P = -\frac{4\ell^2}{2\kappa^2} \lim_{z\rightarrow0} \frac{\delta A_i^P}{z}  = 0
\ee
where we have used~\eqref{Aiis} and that $\de g_{ti} \sim O(z^3)$. 

\subsection{An explicit example}

\begin{figure}[t!]
\centering
\includegraphics[width=8.5cm,clip=true]{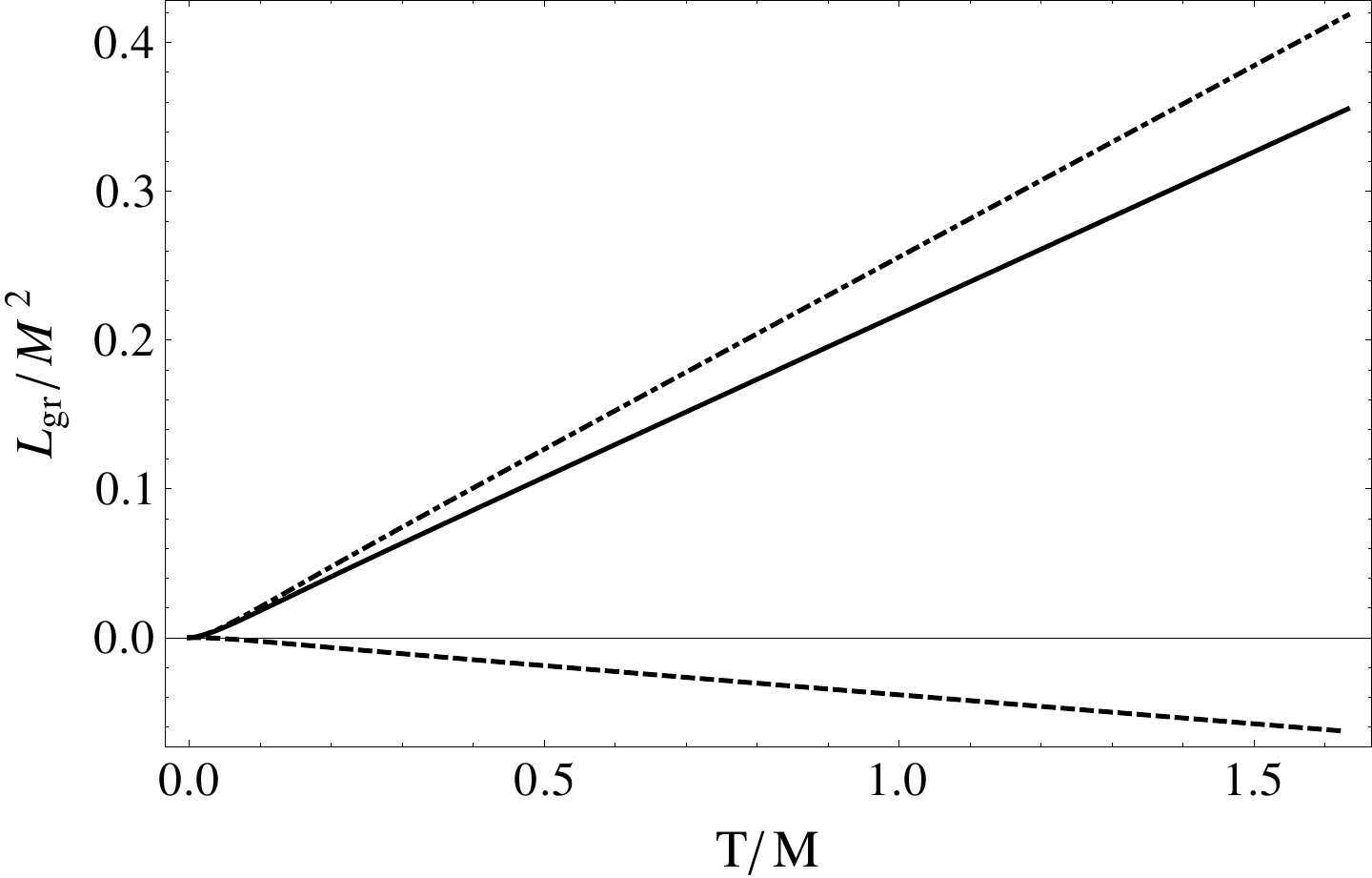}
\caption{Angular momentum density as a function of $T/M$ for gravitational Chern-Simons coupling and non-normalizable scalar field in a quadratic potential with $m^2=-2$, at $\mu=0$. The total angular momentum is represented by solid lines, the first term in Eq. \eqref{l2} by dot-dashed lines and the second term in Eq. \eqref{l2} by dashed lines.}
\label{fig:GR_L}
\end{figure}

We now examine an explicit example. For simplicity we once again consider the setup of Sec. \ref{subsec:ex} with $V (\vt) = \ha m^2 \vt^2$, $m^2=-2$ and $\vt$ non-normalizable with $M$ the scalar source. We exhibit plots of the gravitational angular momentum as a function of $T/M$ in Fig. \ref{fig:GR_L}, with the two terms entering Eq. \eqref{l2} presented separately. We remark that the plots are almost linear, which can be understood from the general structure of Eq. \eqref{l1} as follows: the geometric factor under the integral is roughly proportional to $T^2$ to leading order, while the scalar field is proportional to $M/T$ at leading order, making the overall leading order dependence $L_\mrm{gr} \propto MT$. 

\section{Relation to the chiral magnetic effect and the axial magnetic effect}
\label{sectionIV}

When the scalar field is marginal it is possible to relate our results to the chiral magnetic 
effect and to the axial magnetic effect in $3+1$ dimensions \cite{axialone,axialtwo,axialthree,chiralone,SonSurowka,landsteiner}  via dimensional reduction, as we now explain. 

In $3+1$ dimensions, the gauge anomaly,
\be
\partial_\alpha j^\alpha  = \frac{b_\mrm{CS}}{4} \epsilon^{\alpha\beta\gamma\delta} F_{\alpha\beta} F_{\gamma\delta}, 
\label{fourdanomaly}
\ee
is known to cause
spontaneous generation of the corresponding current,
\be
  j^i = b_\mrm{CS} \mu \epsilon^{ijk} F_{jk}, 
\label{chiralmagnetic}
\ee
and of the momentum density, 
\be
T^{0i} =\frac{b_\mrm{CS}}{2} \mu^2 \epsilon^{ijk} F_{jk},
\label{axialmagnetic}
\ee
where $i,\ j,\ k=1,\ 2,\ 3$ are spatial directions in $3+1$ dimensions. 
These effects are called the chiral magnetic effect for $j^i$ and the axial magnetic effect for $T^{0i}$. 
(The formulas derived in \cite{SonSurowka} in the Landau frame contain terms in higher powers of $\mu$. 
The formulas in the above are in the laboratory frame \cite{landsteiner}.)

In comparison, the Chern-Simons term in our bulk action in $3+1$ dimensions,
\be
S_{CS} = - \frac{\bCS\ell^2}{2\kappa^2} \intop d^4x \sqrt{-g} \vartheta \SF^{ab} F_{ab},
\ee
gives rise to an anomalous divergence of the current $j^\alpha$ on the boundary in $2+1$ dimensions as
\be 
 \partial_\alpha j^\alpha = \frac{2\beta_{\rm{CS}} \ell^2}{\kappa^2} \epsilon^{\alpha\beta\gamma} \partial_\alpha \vartheta F_{\beta\gamma} , 
\ee
where $F_{\beta\gamma}$ is the background gauge field for the boundary CFT. 
Since it is the dimensional reduction of the chiral anomaly (\ref{fourdanomaly}) in $3+1$ dimensions, 
where the scalar field $\vartheta$ in the bulk is identified with the extra component $\vartheta = A_3$
and $F_{i3} = \partial_i \vartheta$, 
we expect effects corresponding to the chiral magnetic effect (\ref{chiralmagnetic}) and to the axial magnetic effect (\ref{axialmagnetic}) to be
\ba
j^i &=& 2b_\mrm{CS} \mu \epsilon^{ij}\partial_j \vartheta , \nn\\
T^{0i} &=&  b_\mrm{CS} \mu^2 \epsilon^{ij}\partial_j \vartheta,
\label{rederiveone}
\ea
where we should identify $b_\mrm{CS} = \bCS\ell^2/\kappa^2$.

We can also include effects due to the axial-gravitational anomalies. In $3+1$ dimensions, the axial-gravitational anomaly, 
\be
\pd_\alpha j^\alpha = \frac{a_\mrm{CS}}{8\pi^2} \ep^{\gamma\delta \eta \theta} R^{\alpha\beta}{_{\eta \theta}} R_{\beta\alpha \gamma \delta} , 
\ee
is known to generate the momentum current
\be
T^{0i} = \frac{a_\mrm{CS}}{2} T^2  \epsilon^{ijk} F_{jk},
\label{axialmagnetic}
\ee
but not the current $j^\alpha$ itself. The corresponding effect in $2+1$ dimensions should be
\be
T^{0i} =a_\mrm{CS}T^2  \epsilon^{ij} \partial_j \vartheta,
\label{rederivetwo}
\ee
with the identification,  $a_\mrm{CS} = 2\pi^2\aCS\ell^2/\kappa^2$.

The dimensional reduction of the chiral magnetic effect and axial magnetic effect, (\ref{rederiveone}) and (\ref{rederivetwo}),
are in agreement with Eqs. \eqref{edgeone} and \eqref{edgetwo} and
consistent with results in our previous paper \cite{Liu:2012zm},
where the scalar field $\vartheta$ is dual to a marginal operator on the boundary CFT.

The main results in this paper, however, are for $\vartheta$ dual to a relevant operator,
which cannot be obtained by dimensional reduction of a massless gauge field
in $4+1$ dimensions. 
There may be a generalization of the chiral magnetic effect and of the axial magnetic effect in $3+1$ dimensions which would correspond to dimensional oxidation of the effects studied in this
paper, and we leave this possibility for future investigation.


\section*{}
\acknowledgments
We thank S.~S.~Gubser, O.~Saremi and D.~T.~Son for useful discussion, and we are grateful to N. Yunes for collaboration on the early stages of this project. We would like to thank K.~Landsteiner and the referee for
their useful comments on the paper.
H. O. and B. S. are supported in part by U.S. DOE Grant No. DE-FG03-92-ER40701. The work of H. O. is also supported in part by 
a Simons Investigator award from the Simons Foundation, the WPI Initiative of MEXT of Japan, and 
JSPS Grant-in-Aid for Scientific Research No. C-23540285. He also thanks the hospitality of the Aspen Center for Physics and
the National Science Foundation, which supports the Center under Grant No. PHY-1066293, and of the Simons Center for Geometry and Physics. The work of B. S. is supported in part by a Dominic Orr Graduate Fellowship. B. S. would like to thank the hospitality of the Kavli Institute for the Physics and Mathematics of the Universe and of the Yukawa Institute for Theoretical Physics. H. L. is supported in part by funds provided by the U.S. DOE under cooperative research agreement DE-FG0205ER41360 and thanks
the hospitality of Isaac Newton Institute for Mathematical Sciences.

{\it Note added.}---When this paper was almost complete, we received the paper \cite{SonWu},
in which holographic models with nonzero angular momentum and Hall viscosity are discussed. Their models are different from those discussed in this paper.

\appendix

\section{Boundary condition at the horizon} \label{app:horb}

The $zt$ component of the Einstein equations reads
\be
\label{deltazt}
f'(z) \pd_i \delta g_{ti}(z,x^i) - f(z)\pd_i \delta g_{ti}'(z,x^i) = 0
\ee
which can be integrated to give
\be
\pd_i \delta g_{ti}(z,x^i) = f(z)W(x^i).
\ee
Since $f(0)=1$ and we choose $\delta g_{ti}$ to be a normalizable perturbation, we must have $W(x^i)=0$ so we conclude
\be
\label{divvi}
\pd_i \delta g_{ti}(z,x^i) = 0.
\ee
Using the $ii$ component of the background Einstein equations the $ti$ component of the Einstein equations reads
\ba
\label{LHSti}
&+&2zfh^2 \epsilon_{ij} \pd_j \left(\pd_x \delta g_{ty}- \pd_y \delta g_{tx}\right) -2zfh \delta g_{ti}'' \nn\\
&+&\lb zfh' + zhf' +4fh \rb \delta g_{ti}'\nn\\&-& 8z^3 fh A_t' \delta\! A'_i(z,x^i) = 0.
\ea
Using \eqref{divvi} $\epsilon_{ij} \pd_j \left(\pd_x \delta g_{ty}- \pd_y \delta g_{tx}\right) = -\pd^2 \delta g_{ti}$ with $\pd^2=\pd^i\pd_i$ this is
\ba
\label{that}
& & -2zfh^2 \pd^2 \delta g_{ti} +\lb zfh' + zhf' +4fh \rb \delta g_{ti}'
\nn \\
& & -2zfh \delta g_{ti}'' - 8z^3 fh A_t' \delta\! A'_i = 0.
\ea
We now count the divergences in \eqref{that}, using that near the horizon
\ba
\label{hasyis}
h(z) &=& \frac{\mathcal{K}}{f(z)}+K_0+K_1(z-z_0)+\dots,\\
\label{hasyis2}
h'(z) &=& \mathcal{K} \frac{f'(z)}{f^2(z)}+K_1+\dots,
\ea
with $\mathcal{K}$ an arbitrary constant. Since the gauge and scalar fields do not diverge at the horizon we obtain the lhs of the Einstein equations to be
\be
-\frac{1}{2}\pd^2 \delta g_{ti}(z_0,x^i) = 0
\ee
and imposing the boundary condition $\delta g_{ti}(z_0,x^i)\rightarrow 0$ at spatial infinity we conclude
\be
\label{dgticondh}
\delta g_{ti}(z_0,x^i) = 0.
\ee

\section{Regularization and renormalization}
\label{sec:app_b}

Consider the action 
\ba
S &=& \frac{1}{2\kappa^2}\intop d^4x \sqrt{-g} \bigg[R - \frac{1}{2} G^{IJ}(\vartheta^K)\pd_a \vartheta^I \pd^a \vartheta^J \cr &-& V(\vartheta^K) - \ell^2 Z^{PQ}(\vartheta^K) F^P_{ab}F^{Qab} + S_{cs} \bigg],
\ea
where either 
\be
S_{cs} = - \ell^2 \bCS C^{PQ}(\vartheta^K) \SF^{Pab} F^Q_{ab}
\ee
or
\be
S_{cs} = - \frac{\aCS\ell^2}{4} \vartheta^{I=0} \SR R.
\ee
Note the gravitational Chern-Simons term can always be written in this form via field redefinition.

{\it A priori}, there are four possible contributions that need to be accounted for: the usual Gibbons-Hawking-York boundary term, a term arising from the variation of the (axionic or gravitational) Chern-Simons term $S_{cs}$, potential additional terms that must be added for the Dirichlet boundary-value problem to be well-defined and local counterterms (see the Appendix of \cite{Liu:2012zm} for details). Thus, we can write
\be
T^\mathrm{bdy}_{\alpha \beta} = \frac{1}{2\kappa^2}\left(2K_{\alpha \beta} - 2h_{\alpha \beta}K + T_{\alpha \beta}^\mathrm{cs} + T_{\alpha \beta}^\mathrm{reg} - T_{\alpha \beta}^\mathrm{ct} \right).
\ee
The CFT stress-energy tensor is obtained by computing the boundary stress-energy tensor $T^\mathrm{bdy}_{\alpha \beta}$ on a plane at finite $z$ parallel to the boundary, multiplying by an appropriate power of $z$ ($z^{-1}$ in our case for the stress-energy tensor with both indices down) and taking the $z\rightarrow 0$ limit, according to the standard AdS/CFT dictionary (see e.g. \cite{KB1999, BFS2002, dHSS2001}).

Let us first concentrate on possible Chern-Simons and regularization contributions to the boundary stress-energy tensor. As explained in the Appendix of \cite{Liu:2012zm}, the gravitational Chern-Simons term does not contribute to the boundary stress-energy tensor and also does not require additional regularization terms. Similarly, the axionic Chern-Simons term is topological, so under the variation we consider it will not contribute to the boundary stress-energy tensor, nor will it require regularization terms. 

We are thus left to analyze possible counterterms. For planar boundaries there is a standard counterterm obtained by adding a cosmological constant term on the boundary, which does not depend on the presence of scalar fields. In addition, there can be scalar-field dependent counterterms, which we can schematically write by adding
\be
\sqrt{-h}H(\vartheta^I)
\ee
to the action, with $H$ some function and $h_{ab}$ the induced metric,
\be
h_{ab} = g_{ab} - n_a n_b, \qquad n_a = \frac{1}{g^{zz}}\delta_a^z.
\ee
The scalar field counterterms contribute 
\be
T^{\mrm{ct},\vartheta}_{ti} \sim H(\vartheta^I)h_{ti} 
\ee
to the $ti$ component of $T^\mathrm{bdy}_{\alpha \beta}$. However, since we are considering the metric perturbations to be normalizable
\be
h_{ti} \sim \mathcal{O}(z) 
\ee
near the boundary. Furthermore, $H(\vartheta^I)$ cannot contain marginal scalar fields, so it must consist entirely of scalar fields decaying as some positive power of $z$ towards the boundary. Since the counterterms must vanish in the absence of any scalar field $H(\vartheta^I)$ must be proportional to at least one positive power of $\vartheta^I$, which introduces at least one more positive power of $z$ in $T^{\mrm{ct},\vartheta}_{ti}$. Thus
\be
T^{\mrm{ct},\vartheta}_{ti} \sim \mathcal{O}(z^{1+\gamma}), \ \gamma>0
\ee
and the scalar field counterterms decay at least one power of $z^\gamma$ too fast near the boundary to contribute to the CFT stress-energy tensor.

We are thus left with the usual boundary stress-energy tensor in the $ti$ component,
\be
T_{ti}^\mathrm{bdy} = \frac{1}{\kappa^2}\left(K_{ti}-h_{ti}K - \frac{2}{\ell}h_{ti} \right).
\ee


\end{document}